\documentstyle[aps]{revtex}

\def\half{{\textstyle{1\over2}}}
\def\lh{{{il}\over{2}}}
\def\upz{uz/\pi}
\def\ups{u^2}
\def\p1half{{\textstyle{{{p+1}\over{2}}}}}
\def\phalf{{\textstyle{{{p}\over{2}}}}}
\def\23phalf{{\textstyle{{{23-p}\over{2}}}}}
\def\twelfth{{\textstyle{{{1}\over{12}}}}}

\input{psfig} 
\begin{document}
\draft
\preprint{PSU-TH-220}
\title{Pair Correlation Function of Wilson Loops
}
\author{Shyamoli Chaudhuri, Yujun Chen, and Eric Novak
\footnote{shyamoli@phys.psu.edu,ychen@phys.psu.edu,novak@phys.psu.edu}}
\address{Physics Department \\
Penn State University \\ 
University Park, PA 16802}
\date{\today}
\maketitle

\begin{abstract}
We give a path integral prescription for the pair correlation function 
of Wilson loops lying in the worldvolume of Dbranes in the bosonic 
open and closed string theory. The results can be applied both in 
ordinary flat spacetime in the critical dimension $d$ or 
in the presence of a generic 
background for the Liouville field. We compute the 
potential between heavy nonrelativistic sources in an abelian gauge 
theory in relative collinear 
motion with velocity $v$$=$${\rm tanh} (u)$, probing length scales down to 
$r_{\rm min.}^2$$=$$2\pi \alpha^{\prime} u$. We predict a universal 
$-(d-2)/r$ static interaction at short distances.
We show that the velocity dependent corrections to the short 
distance potential in the bosonic string 
take the form of an infinite power series in the dimensionless 
variables $z$$=$$r_{\rm min.}^2/r^2$, $uz/\pi$, and $u^2$. 
\end{abstract}

\pacs{PACS numbers:  11.25.-w, 12.38.Aw, 31.15.Kb}


\section{Introduction}
\label{sec:intro}

It is an old result due to Wilson \cite{wilson} that the expectation
value of the loop observable $W({\cal C})$ in QCD has an effective 
description at low energies as a sum over random surfaces with 
fixed boundary:
\begin{equation}
<W({\cal C})> ~ \equiv ~ \left< {\rm tr} ~  P~ {\rm exp} (i 
   \oint_{\cal C} dX \cdot A ) \right> ~
\sim ~ Z_{\rm string}[{\cal C}] \quad .
\label{eq:loopo}
\end{equation}
${\cal C}$ is the closed world-line of a heavy quark--antiquark pair,
the brackets $< \cdots >$ denote the averaging over gauge field 
configurations, and the \protect{quarks} are treated as semi-classical 
sources in the gauge theory \cite{wilson}.
$Z_{string}[{\cal C}]$ denotes the sum over world-sheets terminating
on a fixed curve ${\cal C}$ in spacetime of some string theory: an
open string amplitude with a macroscopic
hole in the world-sheet. In the gauge theory,
$<W({\cal C})>$$\simeq$$e^{-S_{\rm eff.}[{\cal C}]}$ 
is characterized
by an area dependence in the effective action for large loops with
widely separated quark and antiquark world-lines, with crossover to a
perimeter growth of the effective action for large loops with closely
separated world-lines at fixed spatial separation \cite{wilson}. This
behavior is characteristic of the long distance effective dynamics of
a large class of phenomenological string theories including the Nambu-Goto
and Eguchi-Schild strings \cite{wilson,nambugoto,eguchi}.
Dbrane backgrounds of critical string theory in flat spacetime and at
weak coupling \cite{dbrane} enable a universal and quantitative prediction
of the {\em short distance} dynamics of Wilson loops---a regime of QCD that
remains largely unexplored by either analytic or lattice techniques.
Our result is extracted directly from a covariant path integral computation
for the critical Polyakov string with boundaries \cite{polyakov,alvarez},
extending techniques developed in the earlier works
\cite{poltorus}\cite{cohen}. The key ingredient which enables this
prediction is its relationship to the vacuum
energy computation in string theory: unlike in
quantum field theories, the
one-loop cosmological constant in critical string theory can be unambiguously
normalized, an observation due to Polchinski \cite{poltorus}.

Let us review some aspects of the equivalence in Eq.\ (\ref{eq:loopo}) as
known from previous work. The leading quantum correction to the
area law from the long distance effective dynamics of the string theory yields
an attractive and model independent $1/r$ term in the static potential
between two heavy sources in the gauge theory. The static
coefficient is coupling constant and 
cutoff independent, and was first discovered using functional
methods by L\"{u}scher, Symanzik, and Weisz \cite{luscheret} in a semi-classical
quantization of the Eguchi-Schild string. A model independent argument based
on an effective field theory governing the quantum dynamics at large distance
scales of a thin flux tube linking the two sources gave the result 
\cite{luscher}:
\begin{equation}
V(r) = \alpha r + \beta -{{\pi}\over{24}}(d-2) \cdot{{1}\over {r}} + O(1/r^2)
\quad ,
\label{eq:eftresult}
\end{equation}
where $\alpha$, $\beta$ are model-dependent nonuniversal coefficients,
and $d$ is the number of spacetime degrees of freedom of the collective
coordinate for the thin flux tube. The linear term, signaling confinement,
dominates at large separations. The qualitative form of the static potential
in Eq.\ (\ref{eq:eftresult}) has been extensively confirmed in lattice gauge
theory analyses where a high precision measurement of the linear term is
easily performed. Recent work in string theory 
\cite{largen} has examined the long distance
effective dynamics of Wilson loops in certain large N gauge theories using a
conjectured dual description as an effective limit of the type IIB string
theory in AdS spacetimes \cite{malda}. In this limit, gravity decouples 
from the gauge theory on the Dbranes: for
large N, there is a clear separation of scales between the gauge theory 
with effective couplings of $O(g_s N)$ and supergravity with couplings of 
$O(g_s)$. The leading gravitational corrections to the one-loop amplitude 
we consider are of $O(g_s^2)$, naturally suppressed at short distances 
and at weak coupling, even at finite $N$. On the other hand, the long 
distance gauge dynamics of Wilson loops cannot be directly explored by open 
and closed string theory without taking an appropriate large N limit. 
At long distances, the bosonic pair correlation function we derive 
is instead to be interpreted in terms of a different low energy 
theory---gravity. Duality alters this interpretation considerably 
in nonperturbative String/M theory. We defer that discussion to a 
consideration of the superstring \cite{typeI}.

The idea of using the Polyakov string path integral to extend the analytic
estimates for the leading quantum corrections to the static potential at long
distance into the {\em short distance} regime is due to Alvarez \cite{alvarez}.
The implications of a $1/r$ term in the short distance potential between
sources in a nonabelian gauge theory are discussed in \cite{peskin}. At 
short distances, the notion of a thin flux tube no longer holds 
but a direct computation of the short distance potential between
sources can be performed in a renormalizable 
string theory with boundaries. Unlike
the Nambu-Goto and Eguchi-Schild strings, the quantization of the 
bosonic Brink-Di Vecchia-Howe-Deser-Zumino
action \cite{bdhdz} due to Polyakov \cite{polyakov}
treats the world-sheet metric as an independent dynamical field. 
The action is renormalizable,
enabling in principle a closed form analysis of the string functional integral
without the need to take an effective long distance limit. We will carry out
that analysis in this paper for {\em critical} string theory.
We show that Dbrane backgrounds of open and
closed string theory in flat spacetime and at weak coupling \cite{dbrane} provide
a calculable framework within which the short distance behavior of the static
potential can be obtained directly from the Weyl invariant 
string path integral in the critical spacetime dimension. 
The results can also be adapted to generic conformally invariant backgrounds 
of string theory with $c_m$ matter fields coupled to a Liouville field with
total central charge equal to the critical dimension, following \cite{ddk}. 
The loops are taken to lie in the $p$$+$$1$
dimensional worldvolume of a Dpbrane---a hypersurface in a higher dimensional
spacetime on which the gauge fields live. In the nonabelian case
each loop lives in an $N$ index Chan-Paton state of an open and closed string
theory within the worldvolume of $N$ coincident 
Dpbranes. Note that the {\em short 
distance} potential between semi-classical sources lying in the worldvolume of a 
single Dbrane, or coincident Dbranes, is independent of the nonabelian nature 
of the gauge theory if we neglect interactions: for free strings, the colorless 
amplitude will simply scale as $N^2$ for 
$N$ coincident Dbranes. In what follows, we derive an expression for the
spatial correlation function of a pair of Wilson loops lying within the worldvolume
of a single Dbrane in a generic flat spacetime background of the bosonic open
and closed string theory. We extract from this expression a prediction for the
short distance potential between slow moving sources in the worldvolume gauge 
theory with small relative velocity, $v$$=$${\rm tanh} u$$\simeq$$u$, 
and relative position, $r$. This gives: 
\begin{equation}
V(r,u) = - (d-2) \cdot {{1}\over{r}} \left [ 1+ O(z^2) + O(\upz)
+ O(\ups) \right ] \quad . 
\label{eq:strgresult}
\end{equation}
The subleading terms in the potential will 
be obtained in a systematic expansion for small velocities and 
short distances valid down to distances
of order $r_{\rm min.}^2$$=$$2\pi\alpha^{\prime} u$. 
They are succinctly expressed as a convergent power 
series in powers of the dimensionless variables, $z$$=$$r^2_{\rm min.}/r^2$, 
$\upz$, and $\ups$. 
Note that the potential is a quantum effect accounting for the 
fluctuations about the minimum action surface which 
determines the saddle point of the string path integral. It is time reversal 
invariant \cite{polchinskibook}: the power series only contains even powers
of $u$. The numerical coefficient of the static term 
is a measure of the number of degrees of freedom in the theory describing 
the short distance dynamics. As with the L\"{u}scher term in the long distance 
potential between sources in a gauge theory, the static coefficient is 
free of both string coupling constant and string cutoff dependence. 
In a theory with supersymmetry, the leading static term will be absent 
but there is a corresponding velocity dependent potential \cite{typeI}.

In this paper, we will give a prescription for the pair correlation
function of macroscopic loop observables, $M(C_i)$, $M(C_f)$, in bosonic
open and closed string theory using a covariant path integral technique
for one-loop string amplitudes developed by Polchinski \cite{poltorus},
which determines unambiguously their normalization. This technique was
applied to the covariant path integral for off-shell closed
strings---amplitudes with macroscopic holes in the world-sheet mapped to
fixed curves in spacetime, in \cite{cohen}. Explicit
results were obtained for pointlike boundary states but an implementation of
boundary reparameterization for finite sized loops directly in the path
integral has been lacking so far.  We note that there exists a BRST 
analysis of boundary reparameterization invariance (see, for example, 
\cite{birmingham}) but BRST methods are unsuitable for addressing 
issues related to the normalization of the vacuum amplitude. 
The path integral implementation of boundary reparameterization invariance 
we will give is adapted from the work of
Cohen et al \cite{cohen}, and also from the earlier works 
\cite{polyakov,alvarez,poltorus}.
We define the pair correlation function of macroscopic loop observables
as the covariant string path integral:
\begin{equation}
< M(C_i) M(C_f)> ~ \equiv ~
 \int_{[C_i,C_f]} {{[d g][d X ]}\over{{\rm Vol}({gauge})}}
     e^{-S[X,g_{ab};\mu_0,\lambda_0^{(i,f)}]} \quad ,
\label{eq:path}
\end{equation}
a reparameterization invariant sum over world-sheets of cylindrical topology
terminating on fixed boundary curves, $C_i$, $C_f$, which are taken to lie
in the worldvolume of a Dpbrane in flat spacetime. We will gauge both 
world-sheet diffeomorphisms and Weyl transformations of the world-sheet 
metric. Our results are derived for string theory in $d$$=$$26$ critical 
spacetime dimensions although they could be adapted to generic conformally 
invariant backgrounds of string theory with $c_m$ matter fields coupled 
to a Liouville field, with total central charge equal to the critical 
dimension \cite{ddk}, as outlined in section IIIB. $S[X,g_{ab}]$ is the 
bosonic Brink-Di Vecchia-Howe-Deser-Zumino action \cite{bdhdz}
plus appropriate bulk and boundary terms as necessary
to preserve diffeomorphism and Weyl invariance. We compute
quantum fluctuations about a saddle point describing a surface of minimum
action stretched between coplanar loops of fixed length, $L_i$, $L_f$, with
spatial separation, $R$. The short distance potential 
between sources is extracted from the long loop length limit of this
amplitude: $L_i$, $L_f$$\to$$\infty$ with $R$ held {\em fixed}.

It should be noted that the one-loop amplitude with macroscopic boundaries
is free of any coupling constant dependence. Corrections to the leading
term in string perturbation theory are $O(g_{\rm open})$ from splitting,
and $O(g_s)$ and higher order from open and closed string loops. Here
$g_{\rm s}$$=$$g^2_{\rm open}$, where $g_{\rm open}$ is identified with the
Yang-Mills coupling, and gravitational corrections enter at 
$O(g_s^2)$ as closed string loops---suppressed at weak coupling.
Nevertheless, even in Dpbrane backgrounds where the gauge fields live in
$p$$<$$d$$-$$1$ spatial dimensions, evidence 
for the higher dimensional string theory
in which the gauge fields live is present in the numerical coefficient of the
leading term in the short distance potential. The reason is that the
world-sheet fluctuates in all of the spacetime dimensions---both parallel 
and transverse to the worldvolume of the Dbrane. From the viewpoint of a
nonabelian gauge theory, the transverse fluctuations arise from scalar 
fields in the adjoint representation of the gauge group.

We begin in section II with a discussion of classical boundary reparameterization 
invariance, explaining its implementation in the string path integral. A 
boundary state in the bosonic string is specified
by an embedding and an einbein. For fixed embedding of the loops, we give a
boundary diffeomorphism invariant 
prescription for the measure in the path integral,
summing over reparameterizations of the boundary. 
The gauge fixed path integral is derived in detail in section IIIA. 
We determine
the normalization of the 
path integral as in Polchinski's analysis of the torus amplitude in the
bosonic string theory
\cite{poltorus}, extended to string amplitudes with boundaries 
\cite{alvarez,cohen,chaupath}. For completeness, we retain the Liouville dynamics
although our main interest is in string theory in the critical spacetime 
dimension. 
Section IIIB is an aside explaining how this analysis can be 
applied to Polyakov strings with
generic conformally invariant backgrounds for the Liouville field following
\cite{ddk}. Readers whose main interest is in the potential calculation for 
string theory in the critical spacetime dimension can skip this subsection. 
The modifications to the pair correlation function for generic boundary 
conditions pertaining to 
slow-moving sources in relative motion within the 
worldvolume of the Dbrane is derived in 
section IIIC, an analysis similar to the scattering of slow moving
Dbranes in the bulk transverse space \cite{bachas,dkps,polchinskibook}.

The computation of the potential between slow moving sources at short 
distances is given in section IV. We consider heavy sources in the gauge 
theory in relative collinear motion with 
$r^2$$=$$R^2$$+$$v^2 \tau^2$, $v$$<<$$1$, thus giving a simple realization of 
coplanar loops while mimicking nonrelativistic straight line trajectories
in the Euclideanized $X^0$, $X^p$ plane. Here $r$ is their relative 
position, and $\tau$ is the zero mode of the Euclideanized time coordinate, 
$X^0$. We will compute the Minkowskian potential for 
two sources in relative collinear motion with nonrelativistic velocity 
$v$$<<$$1$ for small separations $r$. 
In section IVA, we extract the short distance potential 
between two point sources traversing closed curves in spacetime for
small separations $r$
from the large loop length limit of the pair
correlation function of Wilson loops. The scattering plane $X^0$,$X^p$,
can be wrapped into a spacetime cylinder by periodically identifying the 
coordinate $X^0$. Then the closed world-lines of sources are loops singly 
wound about this cylinder, where the relative position of the sources at 
proper time $\tau$ is $r(\tau)$. We identify these closed world lines with
Wilson loops. Define the effective potential as follows:
\begin{equation}
< M(C_i) M(C_f) > = -i \lim_{T\to\infty} \int_{-T}^{+T} 
    d\tau V_{\rm eff.}[r(\tau),u] 
\quad ,
\label{eq:potenti}
\end{equation}
where we have taken the large loop length limit: 
$L_i$$\simeq$$L_f$$\simeq$$T$$\to$$\infty$, with $R$ held fixed.
The dominant contribution to the potential between sources at 
short distances is from the massless modes
in the open string spectrum. Restricting to these modes,
we can express the potential as a double expansion in 
small velocities and short distances \cite{dkps} with the result:
\begin{equation}
V(r,u) =  - {{{\rm tanh}(u)/u }\over{r(1+\upz)^{1/2}}} 
   \left \{ (d-2) 
{{{\mathbf{\gamma}}(\half, (\pi+ uz)/z )}\over{\Gamma (\half)}}
  + O(z^{2}/(1+\upz)^{2}) \right \} \quad , 
\label{eq:strgresultrect}
\end{equation}
where $z$ is a dimensionless scale factor, $z$$=$$r^2_{\rm min.}/r^2$,
and $r_{\rm min.}$ is the minimum distance that can be probed in 
the small velocity expansion at short distances \cite{bachas}:
$r^2_{\rm min.}$$=$$2 \pi \alpha^{\prime} u$. 
${\mathbf{\gamma}}(\half, (\pi+uz)/z )$ is the incomplete gamma function.
The resummation and systematics of the small velocity 
expansion are discussed in Section IVA. 
The scale factor $z$ determines the magnitude of the velocity dependent
corrections and, therefore, the accuracy of the expansion.  For a given 
accuracy, with fixed $z$ value, we can probe arbitrarily short distances 
$r$ by simultaneously adjusting the velocity $u$. Self-consistency of the
double expansion implies, however, an upper bound on the relative velocity, 
$u$$\le$$u_+$, thereby determining the regime of validity for the small 
velocity approximation. It should be noted that the leading terms in the 
potential can be obtained without use of the 
small velocity expansion. The potential is universal: independent of the 
dimensionality of the higher dimensional Dbrane, the geometrical parameters 
of the loop configuration, and the string scale cutoff. 
We note that there is no evidence for a departure from analyticity in the 
form of the potential between point sources in the bosonic string at short 
distances. The phase transition found in the large $d$ analysis of a class 
of phenomenological string models including the Nambu-Goto string 
\cite{alvarezstatic} appears to be a large $d$ artifact.

D0branes are point-like spacetime topological defects present in the generic 
background of the open and closed 
bosonic string theory. In section IVB, we note that the 
short distance potential between two static D0branes in bosonic string theory 
gives a linear interaction, 
$V_{\rm D0brane}$$=$$-(d-2)r/2\pi\alpha^{\prime}$. The static potential is the
shift in the vacuum energy due to a constant background electromagnetic potential,
but with vanishing electric field strength \cite{dbrane,bachas,polchinskibook}. The 
D0branes are assumed
to have fixed spatial separation in the direction $X^{d-1}$, and to be in 
relative motion with nonrelativistic velocity $v$ in an orthogonal direction
$X^d$ \cite{bachas,dkps,polchinskibook}. The systematics of the
small velocity short distance double expansion, and the value for the minimum
distance probed in the scattering of D0branes, is identical to the results 
obtained in section IVA in general agreement with previous work.

The bosonic string has a tachyon, formally suppressed in obtaining
this result, which must be stabilized in order to obtain a theory with a 
sensible ground state (see, for example, the recent discussion in \cite{sen}). 
Alternatively, it can be eliminated
from the spectrum of physical states, as is possible in 
the fermionic type I and type II string theories \cite{polchinskibook}. 
Evidence for distance scales in String/M theory shorter than the string scale 
down to the eleven dimensional Planck length was originally observed in the 
form of the velocity dependent potential between D0branes in relative motion in
tachyon-free backgrounds of the type II string theory
\cite{bachas,dkps,polchinskibook}.  Dbranes correspond to BPS states in 
the type II supergravity-Yang-Mills theory, solitons with masses of 
$O({{1}\over{g}})$. The observation that solitons with masses of order $1/g$ 
can probe shorter distance scales than ordinary field theory solitons is 
originally due to Shenker \cite{shenker}.
Our result illustrates this principle directly in the gauge theory on the 
worldvolume of a Dbrane in the bosonic string. Stated in complete generality 
for open and closed 
string theories at weak coupling: a Dirichlet boundary, or Wilson loop, can 
probe distance scales arbitrarily shorter than the string scale,
whether in the worldvolume of the Dbrane or in the bulk space orthogonal 
to the brane. We conclude with a brief discussion of 
the implications of our result in the broader context of gauge theory in 
generic backgrounds of String/M theory.

\section{Boundary Reparameterization Invariance}
\label{sec:boundary}

Following Cohen et al \cite{cohen}, the tree correlation function for a pair of 
macroscopic string loops can be represented as a path integral over 
embeddings and metrics on world-sheets of cylindrical topology terminating 
on fixed curves, $C_i$, $C_f$, which lie within the worldvolume of a 
Dbrane: 
\begin{equation}
< M(C_i) M(C_f)> ~=~ 
 \int_{[C_i,C_f]} {{[d g][d X ]}\over{Vol[{\rm gauge}]}} 
     e^{- S_P [X,g_{ab}] - \mu_0 \int d^2 \sigma {\sqrt{g}} } \quad ,
\label{eq:pathi}
\end{equation}
where $S_P$ is the bosonic Brink-Di Vecchia-Howe-Deser-Zumino
action \cite{bdhdz} on a surface with boundaries
terminating on fixed curves. Note that the 
amplitude is free of the string coupling, since
the Euler characteristic, $\chi$, equals zero,
and the boundary cosmological constant terms have
been eliminated in favor of the bulk term since these are not 
independent Lagrange multipliers on a surface of cylindrical 
topology. In this section, we discuss the boundary conditions 
on the world-sheet fields which determine the saddle point of 
the path integral about which we are to compute quantum 
fluctuations.

Begin by considering the boundary conditions on the embedding
coordinates. Setting the variation of the classical action with
respect to the $X^M$ to zero on the boundary yields:
\begin{equation}
(\delta X^M) {\hat n}^a \partial_a X_{M} |_{\partial M} = 0,  
\quad M = 0, \cdots , d-1 \quad .
\label{eq:variation}
\end{equation} 
We will require boundary reparameterization invariance of the amplitude: 
each point on the physical boundary, ${\cal C}$, is identified with a point 
on the piecewise continuous world-sheet boundary $\partial M$, but {\em only 
upto a boundary reparameterization}. Classically, this is most succinctly
expressed as the modified Dirichlet boundary condition on the embedding 
functions \cite{alvarez}: 
$\delta X^{\mu}|_{\partial M}$$
\propto $$
{\hat t}^a \partial_a X^{\mu}|_{\partial M}$, $\mu$$=$$0$, $\cdots$, $p$, 
and zero Dirichlet boundary conditions, $X^m|_{\partial M}$$=$$0$, 
$m$$=$$p+1$, $\cdots$, $d$$-$$1$ in directions 
orthogonal to the brane volume. We can replace Eq.\ (\ref{eq:variation}) 
with the equivalent condition:
\begin{equation}
{\hat n}^a {\hat t}^b \partial_a X^{\mu} \partial_b X_{\mu} |_{\partial M}
  = 0, \quad  \mu = 0, \cdots , p \quad .
\label{eq:modif}
\end{equation}
Note that upon imposition of the modified Dirichlet boundary 
condition on all $d$ coordinates of a {\em space-filling} Dbrane, the 
intrinsic world-sheet metric, $g_{ab}$, satisfies the same classical 
equation of motion as the embedding metric: 
$\gamma_{ab}$$=$$\partial_a X^{M} \partial_b X^{M}$, 
summing on $M$$=$$0$, $\cdots $, $d$$-$$1$. As a consequence, under the 
mapping of the world-sheet boundary to fixed curves in the worldvolume 
of the Dbrane, classically, the physical length of any closed curve is 
identified with the parameter length of a corresponding hole on the 
string world-sheet.

Let $\sigma^1$ be the circle variable parameterizing any hole
on the worldsheet, and ${\hat e}$$=$${\sqrt{\hat g}}|_{\partial M }$ 
be the einbein on the boundary, with fiducial metric ${\hat g}$.
The metric on an arbitrary surface with cylindrical topology 
can be brought to the fiducial form, 
$ds^2$$=$$l^2(d \sigma^1)^2$$+$$(d \sigma^2)^2$, where 
$0$$\le$$\sigma^1$$\le$$1$, $0$$\le$$\sigma^2$$\le$$1$, and 
the area of the surface equals $l$. 
A reparameterization of the boundary, 
$\Sigma$$\in$${\rm Diff}_{\partial M}$,
is a one-to-one invertible mapping of holes on the world-sheet
into corresponding fixed curves in spacetime: 
\begin{equation}
\Sigma ~[X_{\mu}(\sigma^1) |_{(i,f)} ] ~=~ 
  {\tilde x}^{(i,f)}_{\mu}[f^{(i,f)}(\sigma^1)]
\quad\quad 0 \le \sigma^1 \le 1  \quad ,
\label{eq:repb}
\end{equation}
Thus, the ${\tilde x}^{(i,f)} (\sigma^1 )$ are {\em fiducial} maps 
of the boundaries of the world sheet into the spacetime curves $C_i$,
$C_f$, and the $f^{(i,f)}$ are arbitrary diffeomorphisms of 
$\sigma^1$ parameterizing the corresponding holes on the worldsheet.

The path integral sums over quantum fluctuations about a classical
background determined by an extremum of the action. We look for minimum 
action configurations in the classical phase space of the Polyakov 
string. We separate each $X$ into a piece, ${\bar x}$, which solves 
the classical equation of motion with fiducial metric, ${\hat g}$, 
and assumes the functional form ${\tilde x}^{(i,f)}$ on the boundary, 
and a quantum fluctuation which satisfies the zero Dirichlet condition. 
In directions orthogonal to the worldvolume of the Dbrane, the
${\tilde x}^{(i,f)}$ are identically zero. Expanding the classical 
action in a complete set of modes, 
\begin{equation}
x_n^{(i,f)}=\int_0^1 d \sigma^1 x^{(i,f)}
e^{2\pi i n \sigma^1}, \quad\quad n \in {\rm Z} \quad ,  
\label{eq:modes}
\end{equation}
spanning the classical phase space of boundary
configurations, gives \cite{cohen}:
\begin{equation}
S_P[{\bar x};{\hat g}] = 
{{1}\over{4\pi\alpha^{\prime}}} \sum_{n=-\infty}^{\infty}  
{{2n\pi}\over{{\rm sinh} (2n\pi/l)}}
\left [ (|x_n^i|^2 + |x_n^f|^2){\rm Cosh}(2n\pi/l) -
2 {\rm Re} (x_n^i \cdot x_n^{f*}) \right ]
 \quad .
\label{eq:saddleaction}
\end{equation}
Since our interest is in the large loop length limit---where the
dynamics is hopefully universal and independent of the detailed
geometric parameters of the loops, we make a judicious guess for
minimum action configurations, ${\tilde x}^{(i,f)}$, obtaining 
the saddle point action from eq.\ (\ref{eq:saddleaction}). A
simple case is a pair of circular Wilson loops of uniform 
radius $L/2\pi$ separated by a distance $R$. Align the circles
parallel to the $X^p$, $X^{0}$ plane, with their axis in a 
perpendicular direction. Here $X^0$ is a Euclidean coordinate. 
The minimum area world-sheet is a catenoid \cite{kreyszig}:
\begin{equation}
{\bar x}= (a {\rm Cos}(2\pi\sigma^1), a {\rm Sin}(2\pi \sigma^1), 
h(a)) , \quad\quad a^2 = (X^p)^2 +(X^{0})^2 \quad , 
\label{eq:catenoid}
\end{equation}
with $L^{\prime}/2\pi$$\le$$a$$\le$$L/2\pi$. The radial
parameter, $a$, is related to the height of the catenoid, 
$h(a)$, by the equation,
$a$$=$$(L^{\prime}/2\pi) {\rm Cosh}(2\pi h(u)/L^{\prime})$. 
$L^{\prime}/2\pi$ is the minimum radius of cross-section
for the catenoid. It is straightforward to evaluate the Polyakov
action for this surface. Consider the maps that must be 
included in the sum over reparameterizations of the world-sheet 
boundary for this configuration of loops. In general, this is a 
sophisticated problem in the representation theory of the group
${\rm Diff}(S^1)$. However, in the large loop length limit, the
analysis is rather simple since winding number one maps with no
self-intersections are energetically favored. 

\begin{figure}[htb]
\centerline{ \hbox{
   \psfig{figure=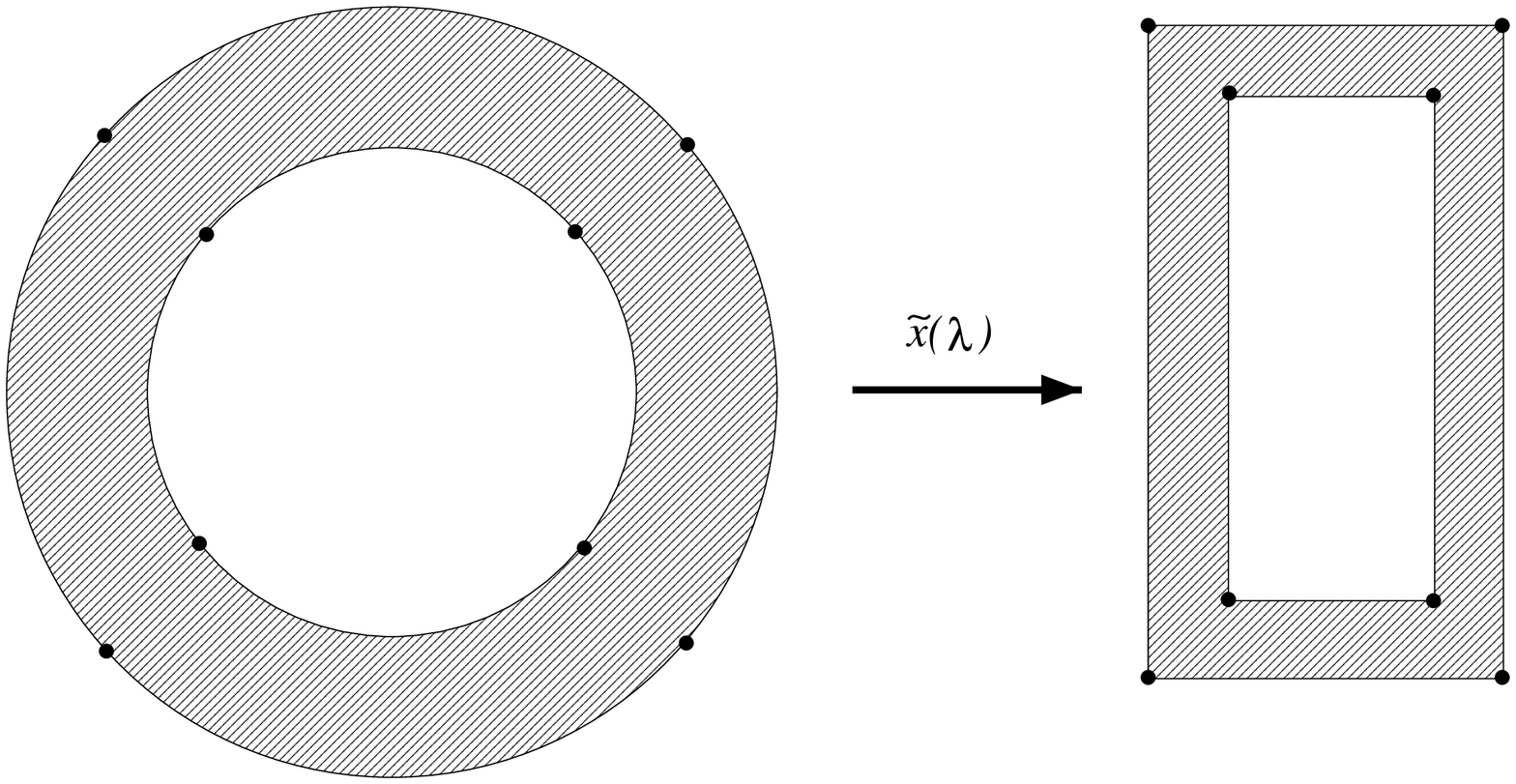,height=2.00in}
 } }
\vspace{0.1in}
\caption{\label{fig:loop}
${\tilde x}(\lambda)$ is a map from the boundaries of the annulus to
a pair of coplanar nested rectangular loops which lie within the 
worldvolume of a Dbrane.}
\end{figure}

This feature of the large loop length dynamics is straightforwardly
captured by considering the simple problem of summing over 
the reparameterizations of loops with one or more marked points. 
For such maps, the sum over reparameterizations of the boundary
is easily implemented in closed form prior to taking a large 
loop length limit. For notational ease, let $\lambda$ denote the
circle variable, $\sigma^1$. We consider non-intersecting 
curves with the following characteristics: each $C^{(i,f)}$ is 
piecewise smooth with $K$ straight line intervals of equal length, 
$s^{(i,f)}_{\alpha}$, and $K$ turning points, 
or corners, $\lambda^{(i,f)}_{\alpha}$, $\alpha$$=$$1$, $\cdots$, 
$K$. Any curvature on the boundary of the world-sheet, 
if present, is permitted only at the corners. As can be 
seen from the Gauss-Bonnet theorem, this would induce a non-vanishing 
Euler characteristic, $\chi_{\rm corner}$$=$$-\sum_{I=1}^2 
\sum_{\alpha=1}^K \delta^{(i,f)}_{\alpha} /2\pi$, and 
consequently a dependence on the string coupling 
constant in the amplitude. The angle terms arise 
from the delta function in the geodesic curvature 
at the corners. The {\em bulk} curvature is however 
required to be smooth: this implies 
that if we consider loop configurations 
with corners, it is convenient to choose the loops to be {\em coplanar}. 
For rectangular, or right-angled, loops, 
the turning angle, $\delta^{(I)}_{\alpha}$$=$$ \pm \pi/2$, for 
every $I$, $\alpha$. The simplest closed loop with net turning 
angle $2\pi$ is a rectangular loop with four edges, $K$$=$$4$. 
The example shown in Figure \ref{fig:loop} is a pair of coplanar 
nested rectangular loops, with $I$$=$$2$, $K$$=$$4$. The net turning 
angle for {\em both} loops vanishes. Thus, 
$\chi_{\rm corner}$$=$$0$ for $C_i$, $C_f$, and there is no 
dependence on the string coupling
constant in the pair correlation function. It should be noted that, 
in general, the presence of corners would be a violation of boundary 
Weyl invariance giving a correction to the Liouville action \cite{weis}.
{\em Any} pair of coplanar nested loops with arbitrary numbers of 
edges having net turning angle zero gives a Weyl invariant loop 
configuration with a well-defined saddle configuration: the stretched
world-sheet in the plane containing the loops. Smooth loops in this 
same class are the boundaries of an annulus---a pair of coplanar
nested circular loops:
\begin{equation}
{\bar x} = ((R_0 +R\sigma^2) {\rm Cos}(2\pi\sigma^1) , 
(R_0 +R\sigma^2) {\rm Sin}(2\pi\sigma^1) )  \quad ,
\label{eq:circularsad}
\end{equation}
where $L_i$$=$$2\pi R_0$, 
$L_f$$=2\pi(R_0 +R)$, and $R$ is their separation in the radial
direction. Periodically identifying Euclidean time, $X^0$ is
the angular, and $X^{p}$ the radial, direction. Now take the 
large loop length limit. Comparing with our discussion of the 
cylindrical spacetime in the introduction, the circular loops 
correspond to the closed world-lines of static sources with fixed 
spatial separation, $R$. From eq.\ (\ref{eq:saddleaction}), and
in the limit $l$$\to$$\infty$, we 
obtain the saddle point action, 
$S_P({\bar x},{\hat g})$$=$$-R^2 l/4\pi\alpha^{\prime}$. 
It is easy to verify an identical result for a pair of 
coplanar nested right-angled loops with arbitrary numbers 
of edges.

\section{Macroscopic Loop Correlation Function}
\label{sec:macro}

We now present the derivation of the closed string amplitude linking 
fixed curves, $C_i$, $C_f$, of length $L_i$, $L_f$, in an 
embedding spacetime
with metric $\gamma_{ab}$$=$$\partial_a X^{\mu} \partial_b X_{\mu}$,
and spatial separation $R$. 
We perform the sum over world sheet metrics using an idea taken from
Cohen et al \cite{cohen}. We begin with the integration over all
embeddings, $(X^m,X^{\mu})$, with fixed fiducial bulk metric, and fixed
fiducial einbeins on the parameter boundary. We choose a fiducial metric on 
the worldsheet, $ds^2$$=$${\hat g}_{ab} d \sigma^a d \sigma^b$, with
${\hat e}$$=$${\sqrt{{\hat g}}}$. Next we sum over world sheet metric
deformations that leave fixed the parameterization of the boundary.
Finally, we perform an integral over the \lq\lq boundary data",
$\{e(\lambda;l^{(i,f)}_{\alpha})\}$, summing 
$\alpha$$=$$1$, $\cdots$, $K$, for all $2K$ intervals, restoring 
boundary reparameterization invariance. This last sum is defined as 
follows (see also the related ideas in \cite{durhuus,weis}).

A boundary state is specified by an einbein and an embedding function, 
$(e,{\tilde x}^{\mu})$. The embedding function is specified by our choice
of saddle configuration, but we wish to leave its parameterization 
unfixed. The sum over einbeins implements 
reparameterization invariance on the boundary, and we must divide by the 
volume of the group of boundary diffeomorphisms, ${\rm Diff}_{\partial M}$. 
Thus, we need a reparameterization invariant measure for the path integration 
over einbeins. The unique choice is Polyakov's quadratic form for metric
deformations \cite{polyakov}, restricted to any given boundary interval.
On any interval, a boundary reparameterization 
$\Sigma$$\in$${\rm Diff}_{\partial M}$, acts as: 
\begin{equation}
\Sigma ~[X_{\mu}|_{s^{(i,f)}_{\alpha}} ] ~=~ 
  {\tilde x}^{(i,f)}_{\mu}[f^{(i,f)}_{\alpha}(\lambda)]
\quad  \lambda_{\alpha -1} 
 \le \lambda \le \lambda_{\alpha}, ~ \alpha = 1, \cdots , K,
\label{eq:repbinterval}
\end{equation}
where the $\lambda_{\alpha}$ are points in the range 
$0\le \lambda \le 1$, and $\lambda$ varies continuously with 
$\lambda_0$$=$$0$ identified with $\lambda_K$$=$$1$.
Thus, ${\tilde x}^{(i,f)}_{\mu}(\lambda)$ is the fiducial map of the 
$\alpha$th interval 
on the circle into the $\alpha$th interval of the curve 
$C_{(i,f)}$, and $f^{(i,f)}_{\alpha}(\lambda)$ is the corresponding 
diffeomorphism of $\lambda$ on the $\alpha$th interval of the circle. 
Schematically, the path integration over quantum fluctuations due to
an arbitrary diffeomorphism of the world sheet has been decomposed:
\begin{equation}
{{1}\over{{\rm Order}(D)}} \int \frac{[d \delta X][d \delta g]}{Vol[
         {\rm Diff}]}  ~\to ~
  \int {{[d \delta e_I(\lambda,{\hat e})]}\over{Vol[{\rm Diff}_{\partial M}]}}
          \int_{[{\hat e}]} \frac{[d\delta g]}{Vol[{\rm Diff}_M]}
                      \int_{[{\hat g};{\hat e}]} [d\delta X]
\quad ,
\label{eq:datab}
\end{equation}
where $Vol[{\rm Diff}_M]$ denotes the volume of the group of diffeomorphisms 
vanishing on the boundary, and $Vol[{\rm Diff}_{\partial M}]$ that of the 
group of boundary diffeomorphisms. We divide by the order of the 
subgroup of the disconnected component of the diffeomorphism group, 
${\rm {\tilde D}}$: discrete diffeomorphisms of the world-sheet left 
invariant under the choice of conformal gauge \cite{poltorus}. Thus, a 
factor of two in the denominator corrects for the two-fold invariance 
of the measure of the path integral under the diffeomorphism:
\begin{equation}
\sigma^1 \to -\sigma^1 \quad .
\label{eq:sigmaones}
\end{equation}
This symmetry will be left invariant under the gauge fixing of
reparameterizations of the world-sheet to be described below.
The measure for embeddings and metrics do not individually 
respect Weyl invariance but, in critical string theory,
their combination is Weyl invariant, and we therefore divide through by the 
volume of the Weyl group. In what follows, we will make this gauge fixing 
procedure explicit. 

\subsection{Gauge Fixing Reparameterizations}
\label{sec:gauge}

The gauge fixing of world-sheet metrics and the path integration of metrics 
and embeddings proceeds as in \cite{poltorus}, except that all harmonic
functions on the world-sheet, scalars and vectors, $(X,\eta_a)$, are 
orthogonally decomposed $\Psi$$=$${\bar \psi}$$+$$\psi^{\prime}$. The 
$\psi^{\prime}$ vanish on the boundary, and the $\bar{\psi}$ are harmonic 
functions taking values ${\bar \psi}|_{\partial M}$$=$${\tilde \psi}$ on 
the boundary. The ${\bar x}$ determine the saddle point configuration 
as described in the previous section.
In general, we will allow for fluctuations of the world-sheet fields on 
$\partial M$, subject only to the constraint that they preserve the normal 
direction to the brane, the fixed embedding of the spacetime curves, and 
the smoothness condition at any corners of the boundary, if present. We will 
assume the reader is familiar with reference \cite{poltorus} and simply 
assemble the different contributions to the path integral.

Begin with the integration over embedding functions with fixed fiducial 
world sheet metric. We choose nested coplanar loops with physical lengths,
$L_i$, $L_f$, and fixed spatial separation, $R$. For each of the $d$ 
scalar degrees of freedom, normalizing the path 
integration over harmonic functions vanishing on the boundary as in 
\cite{poltorus,chaupath} gives the result:
\begin{equation} 
      e^{-R^2 l/4\pi\alpha'} ({\rm det}'\Delta_0)^{-d/2} = 
      e^{-R^2 l /4\pi\alpha'} [\eta(\lh )]^{-d}  \quad ,
\label{eq:embed}
\end{equation}
where the determinant of the Laplacian on scalars is computed 
with the zero Dirichlet boundary condition, and $l$ is the cylinder
modulus defined by the fiducial metric, $ds^2$$=$$l^2(d\sigma^1)^2
$$+$$(d\sigma^2)^2$, $0$$\le$$\sigma^1$$\le$$1$, $0\le$$\sigma^2$$\le$$1$. 
With this choice, the area of the world-sheet is normalized to $l$.

Next consider the integration over metric deformations vanishing on the 
boundary. As in \cite{polyakov,alvarez,poltorus,chaupath}, we isolate the 
dependence on symmetric traceless variations of the metric and divide out by 
the volume of the gauge group $[{\rm Diff}_M]_0$, diffeomorphisms of the 
world-sheet continuously 
connected to the identity and vanishing on the boundary. Normalize the path
integrations on the cylinder as in \cite{poltorus,chaupath}:
\begin{equation}
\int [d \delta {\hat g} ] e^{-|\delta {\hat g}|^2/2} \equiv 
 \prod_{\sigma} \int [d \delta {\hat g} ]_{\sigma} e^{-|\delta 
{\hat g}|_{\sigma}^2/2} = 1 = 
 J_M({\cal \phi},{\hat g}) \int [d \delta {\cal \phi} ]_{e^\phi {\hat g}} 
   \int [d \delta \eta]'_{\hat g} \int_0^{\infty}
 d l e^{-|\delta {\hat g}|^2/2} \quad ,
\label{eq:jacobian}
\end{equation}
where $|\delta {\hat g}|^2$ is the quadratic form for metric deformations,
and $J_M({\cal \phi},{\hat g})$ is the Jacobian from the change of variables 
to deformations of, respectively, the Liouville mode, $\delta {\cal \phi}$, 
diffeomorphisms, $\delta \eta_a$, vanishing on the boundary, and the
cylinder modulus, $l$, computed in \cite{poltorus}. 
The basic assumption 
underlying Eq.\ (\ref{eq:jacobian}) is the locality of the measure: the
integral over elements $[d\delta g]$ is a product of integrals over elements
$[d\delta g]_{\sigma}$ at fixed values of the world-sheet coordinate 
$\sigma$. The only reparameterization invariant local counterterm (free 
of derivatives of the world-sheet metric) is of the form, 
$M \int d^2 \sigma {\sqrt {g}} $, which can be absorbed in a 
renormalization of the bulk cosmological constant,
$\mu_0$, present in the bare action given in Eq.\ (\ref{eq:pathi}).
Thus, the Gaussian integral on the left hand side of Eq.\ 
(\ref{eq:jacobian}) can be set to unity at the cost of renormalizing
$\mu_0$ \cite{poltorus}. The same argument applies to 
any of the world-sheet fields, $(\delta \eta^a, \delta \phi, \delta X)$.
The final value of the renormalized bulk cosmological constant, $\mu_R$,
is set to zero at the end of the calculation, giving a manifestly 
Weyl invariant result for the bosonic string theory in the critical
spacetime dimension $d$$=$$26$.

Taking into account the 
contributions of the conformal Killing vector and zero modes of the 
Laplacian on vectors, $\Delta_1$, the infinity from the 
integration over diffeomorphisms $[d \eta ]'_{\hat g}$ vanishing on the 
boundary 
is cancelled against the volume of the gauge group $[{\rm Diff}_M]_0$.
The result is an integral over the cylinder modulus times the quantum 
functional integral for Liouville field theory:
\begin{equation}
J_M(\phi , {\hat g})
     \int [d \delta \phi ]_{e^\phi {\hat g}} ~ e^{-S_L[\phi,{\hat g}] 
  - S_{\rm boundary}[\phi,{\hat e}]}
                 \quad ,
\label{eq:bulkm}
\end{equation}
where $S_{\rm boundary}$ 
includes any boundary terms necessitated by the 
world-sheet gauge symmetries. $S_L[{\cal \phi},{\hat g}]$ is the 
unrenormalized Liouville action \cite{polyakov}:
\begin{equation}
S_L[\phi,{\hat g}] = {{d - 26}\over{48 \pi}}  
\int_M d^2 \sigma {\sqrt {\hat g}} [ \half {\hat g}^{ab} 
   \partial_a \phi \partial_b \phi
+ {\hat R} \phi ] - \mu_0 \int_M d^2 \sigma {\sqrt {\hat g}} e^{\phi}  
\quad ,
\label{eq:liouville}
\end{equation} 
with integration norm given by:
\begin{equation}
|\delta \phi|^2 = \int d^2 \sigma {\sqrt{\hat g}} e^{\phi} (\delta \phi)^2 \quad .
\label{eq:normphi}
\end{equation} 
We will treat the measure for the einbeins following \cite{poltorus,cohen}.
Note that the quantum functional integral in Eq.\ (\ref{eq:bulkm}) denotes
all possible bulk and boundary deformations of the Liouville field. In 
particular, it receives corrections from the measure for diffeomorphisms 
on the boundary as is shown below.

We continue with the sum over metric deformations nonvanishing on the boundary, 
orthogonal to the modes summed in Eq.\ (\ref{eq:jacobian}). The 
metric on the world-sheet is ${\hat g}_{ab}e^{\phi}$, and the fiducial 
einbein induced on the boundary is ${\hat e}$$=$$\sqrt{{\hat g}}$.
The length of either boundary in the cylinder metric,
$\int_0^1 d\lambda {\hat e}$, equals $l$. A variation of the einbein on
the $\alpha$th interval is the result of a diffeomorphism,
$\lambda$$\to$$f^{(i,f)}_{\alpha}(\lambda)$, and a possible shift,
$\delta \phi$, in the Liouville field. Thus, 
\begin{eqnarray}
&&\{ (({\hat e}+\delta e)[\lambda+\delta f^{(i,f)}_{\alpha} (\lambda)])(1+
        {{d}\over{d \lambda}}(\delta f^{(i,f)}_{\alpha} )) +
  {\hat e} \delta \phi \} e^{\phi} \\ \nonumber
 && \quad = \{ {\hat e}(\lambda) + 
{\hat e}\cdot {{d}\over{d \lambda}}\delta f^{(i,f)}_{\alpha} 
  + \delta e(\lambda)
+ O( \delta f^{(i,f)}_{\alpha})^2 + {\hat e} \delta \phi \} e^{\phi} \\ \nonumber
 && \quad = \{ l^{(i,f)}_{\alpha} + 
 \delta \rho^{(i,f)}_{\alpha} [f^{(i,f)}_{\alpha} (\lambda)] \} e^{\phi} \quad , 
\label{eq:einb}
\end{eqnarray}
where $l^{(i,f)}_{\alpha}$ is the length of the $\alpha$th interval of the
corresponding hole on the worldsheet, and $\delta \rho $ is a rescaling of 
the einbein which can always be absorbed in a shift of the Liouville field on 
the boundary. The restriction of the quadratic form for metric deformations to 
the boundary, $\partial M$, gives a measure on the tangent space to the space of 
einbeins on any given interval: 
\begin{equation} |\delta e|^2 = \int d \lambda ({\hat e} (\lambda; l ))^{-1} 
     (\delta e(\lambda;l))^2 
          = \int d \lambda [ - {\hat e}(\lambda) (\delta f) 
             {{d^2}\over{d \lambda^2}} (\delta f) + 
                        {{(\delta \rho [f(\lambda)])^2}\over{{\hat e}}} ] , 
\label{eq:einmeasure}
\end{equation}
where the zero mode, $\delta \rho_0 [f(\lambda)]$, is the functional change 
in the length of the interval induced by a diffeomorphism.\footnote{Note that 
the conformal class of the metric determines the length of any boundary circle,
a modulus of the surface. Thus, rescalings of the fiducial einbein must 
be absorbed in a shift of the Liouville field on the boundary in order that
the conformal class of the metric is left unchanged.} 
Normalizing the path integrals as in Eq.\ (\ref{eq:jacobian}): 
\begin{equation}
1 \equiv \int [d \delta e ] e^{-|\delta e |^2/2 } = 
 J_{\partial M} (\hat e) \int [d \delta f]_{\hat g} [d \delta 
  \rho ]_{{\hat g}e^{\phi}} 
   e^{-|\delta e|^2/2} \quad ,
\label{eq:jacobianbound}
\end{equation}
where the Jacobian $J_{\partial M}$ is obtained as before, from a change of
variables to deformations of boundary diffeomorphisms,
$\delta f$, and einbein rescalings, $\delta \rho $. Since a rescaling of the
einbein is absorbed by a shift of the Liouville field we can, with no loss
of generality, set $\rho $$=$$0$. Consequently, the integration over 
$[d\delta \rho]$ can be consistently 
dropped from the path integral. The infinity from the integration over 
diffeomorphisms will be cancelled by the volume of the 
gauge group of diffeomorphisms on the 
boundary, ${\rm Diff}_{\partial M}$, which has no disconnected part. 
Combining this analysis of boundary deformations with the bulk deformations
in Eq.\ ( \ref{eq:jacobian}), we can write:
\begin{equation}
1 \equiv 
 [{{(A/2\pi)^{1/2} ({\rm det}\chi^{ab}\chi_{ab})^{1/2}
   ( {\rm det}^{\prime} \Delta_1)^{1/2} }
  \over{({\rm det} Q_{ab}
/2\pi )^{1/2} }} ] 
 \prod_{I=1}^2 
({\rm det}^{\prime}[ -{{1}\over{l_I^2}} 
     {{d^2}\over{d \lambda^2}}])^{1/2} 
  \int [d\delta \phi]_{{\hat g}e^{\phi}} 
   e^{-|\delta g|^2/2} .
\label{eq:jacboth}
\end{equation}
The factor in square brackets is 
the Jacobian, $J_M$, derived in \cite{poltorus}.
$A$ is the world-sheet area, $A$$=$$\int d^2 \sigma {\sqrt{g}}$,
and the term in the denominator arises from 
conformal Killing vectors, if present. On the cylinder,
the area, $A$$=$$l$, in the metric defined 
above, and we have a single conformal Killing vector. The
functional determinant of the Laplacian on vectors is 
computed as in \cite{poltorus,cohen}. With the fiducial cylinder
metric given above,
\begin{equation}
J_M =   {{(l/2\pi)^{1/2} \cdot ({{2}\over{l^2}})^{1/2} \cdot (\half
l^{2} \eta^4 ( \lh ))^{1/2}}\over{(l^3/2\pi)^{1/2}}} 
  \quad ,
\label{Jacobm}
\end{equation}
upto $\phi$ dependent terms absorbed in the Liouville action.

The second term in Eq.\ (\ref{eq:jacboth}) is the boundary Jacobian, 
$J_{\partial M}$, the product of independent determinants 
for each of $K$ intervals, $s_{\alpha}^I$, $\alpha$$=$$1$, $\cdots$, 
$K$. Note that the diffeomorphism acts on each circle as a whole, but 
independently on each of the two boundaries. Since the boundaries 
have a common parameter length, $l$, we obtain:
\begin{equation}
J_{\partial M} = {\rm det}^{\prime}[-{{1}\over{l^2}} 
     {{d^2}\over{d \lambda^2}}] = 2l \quad .
\label{eq:jacobb} 
\end{equation}
A similar path integration appears in the problem of obtaining the 
off-shell propagator for a relativistic point particle, a discussion 
of which appears in reference \cite{cohen}. See, also, the ansatz for 
a scalar quark loop given in \cite{durhuus}.

Assembling Eqs.\ (\ref{eq:datab}), (\ref{eq:embed}), (\ref{eq:bulkm}),
(\ref{eq:liouville}), and (\ref{eq:jacboth}), 
our result for the pair correlation function of piecewise smooth 
macroscopic loops, $C_i$, $C_f$, at fixed separation, $R$, is:
\begin{equation}
< M(C_i) M(C_f)> =  
      \int_0^{\infty} 
  dl e^{-R^2 l/4\pi\alpha'} [\eta( \lh)]^{2-d} 
  \int {{ [d\delta \phi]_{{\hat g}e^{\phi}} }\over{ Vol({\rm Weyl}) }}
 ~ e^{ S[\phi;{\hat g}]} ,
\label{eq:pairc}
\end{equation}
where $S[\phi;{\hat g}]$$=$$S_L[\phi;{\hat g}]$$+$$S_{\rm boundary}$
is the action for the Liouville field including boundary terms.
In the critical spacetime dimension, $d$$=$$26$, the Liouville 
dynamics entirely decouples, and we can consistently set 
$\phi$ to zero in Eq.\ (\ref{eq:pairc}) while dividing out by the 
volume of the Weyl group.

\subsection{Generic Liouville Backgrounds} 
\label{sec:liouville}

It is possible to consider the cases $c_{matter}$$<$$25$ following the
method in \cite{ddk}. We require that the path integral 
expression for the loop correlation function
preserve quantum conformal 
invariance. We begin by suppressing quantum fluctuations and restrict to the 
zero mode, $\phi_0$, noting that the classical equation of motion
is that of a free scalar field in the regime $\phi_0$$\to$$-\infty$: 
the exponential potential is suppressed. 
We will preserve this asymptotic property in defining the quantum 
theory: the wavefunctions (operators) of Liouville conformal 
field theory are required to match smoothly to free field states 
in the $\phi_0$$\to$$-\infty$ regime, characterized by momentum and 
occupation number alone.

Quantum Liouville conformal field theory can be defined by a 
functional integral over a renormalized Liouville field, $\phi_R$, 
with conformally invariant free field norm:
\begin{equation}
|\delta \phi_R|^2 = \int d^2 \sigma {\sqrt{\hat g}} (\delta \phi_R)^2 \quad .
\label{eq:normphir}
\end{equation}
The ansatz of \cite{ddk} is that 
the effects of renormalization can be lumped in the potential, leaving 
a kinetic term for $\phi_R$ with the canonical normalization of a 
free scalar field theory. Thus, we write:
\begin{equation}
     \int [d \delta \phi ]_{e^\phi {\hat g}} ~ e^{-S_L[\phi,{\hat g}]
   + S_{\rm boundary}[\phi ,{\hat e}] }
     = \int [d \delta \phi ]^{\prime}_{{\hat g}} \int_{-\infty}^{\infty}
              d\phi_0 ~ e^{-S[\phi_R,{\hat g}]  }  \quad ,
\label{eq:bulkml}
\end{equation}
where, $S[\phi_R]$, includes all possible renormalizable 
terms in the bulk, and on the boundary, that preserve both diffeomorphism 
and Weyl invariance. Note that a corner anomaly is a spontaneous 
breaking of Weyl invariance on the boundary, contributing an additional
term not included in $S[\phi_R,{\hat g}]$.

In a Weyl invariant theory, the renormalized action $S[\phi_R,{\hat g}]$ 
takes the general form:
\begin{eqnarray}
S[\phi_R,{\hat g}] =&& {{1}\over{8 \pi}}  
\int_M d^2 \sigma {\sqrt {\hat g}} [ \half {\hat g}^{ab} 
   \partial_a \phi_R \partial_b \phi_R
+ Q {\hat R} \phi_R ] - \mu_R \int_M d^2 \sigma 
  {\sqrt {\hat g}} e^{\alpha \phi_R}  \nonumber \\
 && \quad - \sum_{I=1}^2 \lambda_R^{(I)} \int_{C_I} d\lambda {\hat e}
     e^{\beta^{(I)} \phi_R} \quad, 
\label{eq:sconf}
\end{eqnarray}
where $Q$, $\alpha$, and $\beta^{(I)}$, are constants determined 
by the requirement \cite{ddk} that every term in Eq.\ (\ref{eq:sconf}) 
be a dimension one primary field, in a conformal field theory of 
vanishing total central charge: $c_m$$+$$c_{\phi}$$+$$c_{\rm ghosts}$$=$$0$. 
The renormalized bulk and boundary
cosmological constants, $\mu_R$, $\lambda_R^{(I)}$ are arbitrary 
marginal couplings in the conformal field theory. 
With no loss of generality, we could set the boundary cosmological
constant term on the cylinder to zero, retaining $\mu_R$.

The only mode of $\phi_R$ that survives on the boundary 
is $\phi_0$, and the modes $\phi_R^{\prime}$ 
satisfy Dirichlet boundary conditions as in 
\cite{alvarez,poltorus,cohen,chaupath}. Then, conformal
invariance requires:
\begin{equation}
Q = {\sqrt{{c_m -25}\over{3}}} , \quad \alpha = ( \pm {\sqrt{c_m-1}}
 - {\sqrt{c_m - 25}})/2 {\sqrt{3}} \quad ,
\label{eq:constants}
\end{equation}
the upper sign matching the dimension of the cosmological constant
operator as computed in a semi-classical $c_m$$\to$$-\infty$
saddle point evaluation of the path integral for the Liouville
field $\phi$ \cite{durhuus,ddk}. We will not pursue these cases
further since our main interest in this paper is string theory in
the critical spacetime dimension which corresponds to the theory 
with $c_m$$=$$25$, $c_{\phi_R}$$=$$1$.

\subsection{Generic Boundary Conditions}
\label{sec:genericbc}
 
Following \cite{bachas,dkps}, it is an easy extension to compute 
the pair correlation function with boundary conditions pertaining 
to closed world lines for a pair of slow moving sources in relative 
motion. Consider a pair of coplanar nested rectangular loops with the 
plane of the loops aligned parallel to the $(X^0,X^p)$ plane.
$X^0$, $X^p$ are both Euclidean coordinates. Let us rotate one of 
the loops relative to the other through an angle $\phi$ in the 
($X^0$,$X^{p}$) plane, and take the large loop length limit:
$L_i$ $\simeq$ $L_f$ $\simeq$ $T$$\to$$\infty$, with $R$ held 
fixed. Upon analytic continuation, $X^0$$\to$$iX^{0}_M$, the loops 
may be interpreted as the closed world-lines of slow moving scalar 
quarks in collinear motion at short distances $r$ and with relative
velocity $v$$=$${\rm tanh}(-i\phi)$$<<$$1$.

Consider the boundary conditions on $X^0$, $X^p$. We leave the boundary 
conditions at one end-point fixed, and
change the condition at the other end-point to: $X^{p}|_{C_f}$$=$$vX^0$.
Parameterize the world sheet with open string endpoints 
$\sigma^2$$=$$0$,$1$ at boundaries, 
$C_i$, $C_f$, respectively, and an open string loop parameterized,
$0$$\le$$\sigma^1$$\le$$1$. Recall that $\sigma^1$ is identified with 
the fiducial circle variable $\lambda$ defined in the previous section. 
A complete set of eigenfunctions of the scalar Laplacian 
is composed from the basis:
\begin{equation}
\psi^{(\alpha)}_{(n_1,n_2)} = e^{2 n_1 \pi i \sigma^1}
  {\rm Sin}([n_2 +\alpha]\pi \sigma^2) \quad ,
\label{eq:modealpha}
\end{equation}
where $\alpha$ takes values: $-iu/\pi$, or $1$$+$$iu/\pi$. The velocity has
been parameterized as $v$$=$${\rm tanh} u$ and we work in the small velocity
approximation with $v$$\simeq$$u$. The remaining $d$$-$$2$ 
embedding coordinates satisfy the zero Dirichlet boundary condition as 
in section IIIA. The functional determinant of $\Delta_0$ takes the form:
\begin{equation}
{\rm det} \Delta_0^{(\alpha)} = \prod_{n_1,n_2}
[  ({{4 \pi^2}\over{l^2}}) (n_1^2 + {{(n_2 + \alpha)^2l^2}\over{4}}) ]
\quad ,
\label{unregdet}
\end{equation}
where $l$ is the cylinder modulus defined above, and 
$-\infty$$\le$$n_1$$\le$$\infty$, $n_2$$\ge$$0$. This is computed
using zeta function regularization as in \cite{poltorus}: 
\begin{equation}
{\rm ln ~ det} \Delta = - \lim_{s \to 0}
    {{d}\over{ds}} \sum_{n_1,n_2} [ 
({{4 \pi^2}\over{l^2}}) (n_1^2 + {{(n_2 + \alpha)^2l^2}\over{4}}) ]^{-s} 
\quad . 
\label{eq:zetadet} 
\end{equation}
The infinite sum over $n_1$ is expressed as a contour integral
by a Sommerfeld-Watson transform. The contour ${\cal C}$ runs 
counterclockwise from $\infty$$+$$i\epsilon$ to 
$-\infty$$+$$i\epsilon$. The result is:
\begin{equation}
\label{determi}
- \lim_{s \to 0} {{d}\over{ds}} ({{4 \pi^2}\over{l^2}})^{-s}  
  \oint_{\cal C} {{dz}\over{2\pi i}} 
\sum_{n_2\ge 0} [z^2 + 
 {{(n_2 + \alpha)^2l^2}\over{4}} ]^{-s} {\rm Cot}(\pi z)  \quad . 
\label{eq:transform}
\end{equation}
Writing the cotangent as $-i {\rm Cot}(\pi z)$$=$$ 2/(1-e^{-2\pi iz}) -1 $, 
we can extract the contribution from the integral that is 
finite in the limit $s$$\to$$0$:
\begin{equation}
\label{harmosc}
2 \sum_{n_2 \ge 0} {\rm ln} | 1-e^{-\pi l(n_2 + \alpha)}|
\quad ,
\label{eq:harmosc}
\end{equation}
The term singular in the limit $s$$\to$$0$ has a finite $l$ dependent
remnant whose coefficient can 
be identified as the regulated vacuum energy of a complex scalar:
\begin{eqnarray}
- \pi l E_0 &&\equiv \lim_{s \to 0} {{d}\over{ds}} ({{4 \pi^2}\over{l^2}})^{-s}  
  \oint_{\cal C} dz \sum_{n_2\ge 0} [z^2+{{l^2(n_2+\alpha)^2}\over{4}} ]^{-s} 
\nonumber\\  
&&= \lim_{s \to 0} {{d}\over{ds}} \{ ({{4 \pi^2}\over{l^2}})^{-s}  
 l^{1-2s} \zeta(2s-1,\alpha) {{(\Gamma(1-s))^2}\over{\Gamma(2-2s)}}
 2^{1-2s} {\rm tan}(\pi s) \} \nonumber \\
&&= -\pi l[(\alpha - \half)^2 - \twelfth ] \quad .  
\label{eq:vacuum}
\end{eqnarray}
Combining Eqs.\ (\ref{eq:harmosc}) and (\ref{eq:vacuum}), and with
$q$$=$$e^{-\pi l }$, gives the result:
\begin{eqnarray}
({\rm det} \Delta^{(v)})^{- 1/2} &&=
    q^{-1/12 + ({{u^2}\over{\pi^2}} -i{{u}\over{\pi}})/2 } \nonumber \\
 &&\quad \times  \prod_{n_2=0}^{\infty}
 [(1 - q^{n_2-iu/\pi})(1-q^{n_2+1+iu/\pi})]^{-1} \quad ,
\label{eq:determs}
\end{eqnarray}
which can be written as the ratio of Jacobi theta functions. Setting 
$u$$=$$0$ and suppressing the $n_2=0$ term in the result recovers 
the cylinder determinant for a pair of real 
scalars with Dirichlet boundary condition: $[\eta( \lh )]^{-2}$.

Substituting in Eq.\ (\ref{eq:pairc}), we obtain an analogous
result for the pair correlation function of macroscopic loops 
in critical string theory with boundary conditions pertaining 
to sources in slow relative motion:
\begin{equation}
< M(C_i) M(C_f)> =  
       2 \int_0^{\infty} 
  dl e^{-R^2 l/2\pi\alpha'} \eta(il)^{-22} 
 {{\eta(il) e^{-u^2l/\pi}}\over{i \Theta_{11}({{ul}\over{\pi}},il)}}
\quad .
\label{eq:paircmotion}
\end{equation}
For convenience, we have rescaled $l$$\to$$2l$ in the integral.

\section{Short Distance Potential Between Sources}
\label{sec:potential}

We will now compute the potential between heavy nonrelativistic sources 
in the gauge theory in relative 
collinear motion with velocity $v$$=$${\rm tanh}u$$\simeq$$u$. Parameterize 
the closed world-lines of the sources by the proper time variable, $\tau$, 
the zero mode of the Euclidean embedding coordinate $X^0$.  Let 
$r(\tau)^2$$=$$R^2$$+$$v^2\tau^2$ denote the relative coordinate of the 
two sources in the $X^{0}$, $X^p$ plane, where $R$ is their static separation. 
We express the amplitude 
as an integral over Minkowskian time, $-i\tau$. 
The loops are identified with the closed world-lines 
of a heavy quark-antiquark pair in the gauge theory. This computation is 
described in section IVA. We emphasize that
the open string theory results derived in section IVA are 
{\em only} to be applied to the short distance limit of the potential
between heavy sources in the gauge theory.

D0branes are point-like spacetime topological defects in the 
bosonic string theory \cite{polchinskibook}. 
Following \cite{bachas,dkps,polchinskibook}, 
in section IVB we compute the short distance interaction between
two D0branes in the bosonic string theory
obtaining a linear {\em repulsive} static interaction.
The systematics of the small velocity and short distance double
expansion yields similar conclusions for the minimum distance 
as in section IVA.

\subsection{Wilson loops and a short distance $1/r$ potential}
\label{sec:oneover}
 
We define the potential between sources traversing fixed worldlines, 
$V_{\rm eff.}[r(\tau),u]$, as follows:
\begin{equation}
< \cdots > = -i \int_{-T/2}^{T/2} d\tau V_{\rm eff.}[r(\tau),u] \quad ,
\label{eq:potent}
\end{equation}
and take the limit $T$$\to$$\infty$, with $r$ held fixed. Then,
\begin{equation}
V_{\rm eff.}(r,u) = -
4 (8 \pi^2 \alpha ')^{-1/2}
  \int_0^{\infty} dl
 e^{-r^2 l/2\pi\alpha' } {l}^{1/2}
 \eta(il)^{-21}
 {{ {\rm tanh}(u) e^{-u^2l/\pi}}\over{ \Theta_{11}({{ul}\over{\pi}},il)}}
\quad .
\label{eq:potential}
\end{equation}
In the limit of short distances, the amplitude is 
dominated by the exchange of the lowest lying modes in the open string 
mass spectrum. We therefore expand in powers of $e^{-2\pi l}$, organizing 
the integrand as an infinite summation over open string modes, and restrict 
to the lowest lying states. We suppress the leading contribution from the 
open string tachyon---absent in any stable background, 
and focus on the subleading contribution from massless open string modes. 
We will show that the short distance potential between sources in the bosonic 
string--- analogous in some respects to a nonsupersymmetric 
background of the superstring, has a static remnant originating in the
massless modes, a measure of the 
degrees of freedom determining the short distance dynamics of 
Wilson loops. Consider, 
\begin{eqnarray}
V_{\rm eff.}(r,v) =&& -
2 (8\pi^2 \alpha ')^{-1/2}
  \int_0^{\infty} dl
 e^{-r^2 l/2\pi\alpha' } {l}^{1/2} {{{\rm tanh}(u)
e^{-u^2l/\pi}}\over{{\rm Sin}(ul)}}
\nonumber \\
&& \quad \quad \times [ e^{2\pi l}
   + (22 + 2{\rm Cos}(2ul)) + O(e^{-2\pi l})] \quad ,
\label{eq:lightopen}
\end{eqnarray}
where the restriction to massless modes gives:
\begin{equation}
V(r,u) = - 2 (8 \pi^2 \alpha')^{-1/2} 
  \int_0^{\infty} dl
 e^{-r^2 l/2\pi\alpha' } {l}^{1/2} {{{\rm tanh}(u) e^{-u^2 l/\pi} }\over{ {\rm Sin}(ul)}}
 [ 22 + 2{\rm Cos}(2ul) ]  \quad .
\label{eq:massless}
\end{equation}
We will now assume small velocities and short distances, performing a 
double expansion in the variables $r$, $u$. The regime of validity for the
small $u$ expansion is determined by the behavior of the cosecant function.
We can perform a Taylor expansion in the first half-period of its 
argument, $0$$\le$$ul$$<$$\pi$. Consider the corrections to this result from
the integration domain $ul$$\ge$$\pi$. The sine function changes sign at every 
$n\pi$, $n$$\in$${\rm Z}^+$, so that the regions, $n\pi$$\pm$$\epsilon$,
where the integrand is singular can be excised from the domain of 
integration. This leaves the intervals:
$n\pi$$+$$\epsilon$$\le$$ul$$\le$$(n+1)\pi$$-$$\epsilon$. For sufficiently 
small $u$ values the oscillations in the integrand will be increasingly 
rapid, smearing out the integral \cite{polchinskibook}.
The result can always be bounded, or
evaluated by numerical integration, as a self-consistency check on the 
validity of the small velocity short distance approximation. This check
provides an upper limit, $u_+$, on the permissible velocities. With this 
restriction, the contribution from the domain $l$$>$$\pi/u_+$ can be 
dropped and we will suppress it in what follows. Upon Taylor expansion
of the periodic functions in the integrand, the potential can therefore be 
written as:
\begin{eqnarray}
V(r,u) =&& - 2 (8 \pi^2 \alpha')^{-1/2} 
  \int_0^{\pi/u_+ } dl
 e^{-r^2 l/2\pi\alpha' } {l}^{-1/2} \cdot e^{-u^2 l/\pi} {\rm tanh}(u)/u 
\nonumber\\
&& \quad \times \left [ 24 + \sum_{k=1}^{\infty} C_k (ul)^{2k} +
 \sum_{k=1}^{\infty} \sum_{m=1}^{\infty} C_{k,m} (ul)^{2(k+m)} \right ] 
 \quad ,
\nonumber \\
\label{eq:double}
\end{eqnarray}
where the coefficients of the expansion in powers of $ul$ take 
the form:
\begin{eqnarray}
C_k =&& {{(-1)^{k} 2^{2k+1}}\over{2k!}} + {{48|B_{2k}|}\over{2k!}}
(2^{2k-1} -1)  \cr
C_{k,m} =&& (-1)^m 2^{2(m+1)} {{(2^{2k-1}-1)|B_{2k}|}\over{2k! 2m!}}
\quad ,
\label{eq:coefficients}
\end{eqnarray}
and the $B_{2k}$ are the Bernoulli numbers.
Integrating over $l$ gives a systematic expansion for
the potential in powers of $u^{2}/r^{4}$. Let us define a 
dimensionless scaling variable, $z$$=$$r^2_{\rm min.}/r^2$,
where $r_{\rm min.}^2$$=$$2\pi\alpha^{\prime} u$. The
velocity dependent corrections to the potential are succinctly
expressed as convergent power series in the dimensionless 
variables, $z$, $\upz$, and $\ups$: 
\begin{eqnarray}
V(r,u) &=& - {{1}\over{\Gamma(\half) }} 
   {{{\rm tanh}(u)/u}\over{r(1+\upz)^{1/2}}} 
\left\{  24 \cdot {\mathbf{\gamma}}(\half,
(\pi+uz)/z) \right.
\nonumber\\
&&+ \sum_{k=1}^{\infty} C_k  
{\mathbf{\gamma}}(2k+\half,(\pi+uz)/z) [ z/(1+\upz)]^{2k} 
\nonumber\\
&&+ \sum_{k=1}^{\infty} \sum_{m=1}^{\infty} \left. C_{k,m} 
{\mathbf{\gamma}}(2(k+m)+\half,(\pi+uz)/z)  [z/(1+\upz)]^{2(k+m)} 
\right\} .
\nonumber\\
\label{eq:gammas}
\end{eqnarray}
The ${\mathbf{\gamma}}(\nu,(\pi+uz)/z)$ are incomplete gamma functions.
In writing Eq.\ (\ref{eq:gammas}) we have assumed $u/u_+$$<<$$1$.
Note that if the variable $z$ is taken to zero, for distances 
$r^2$$>>$$r_{\rm min.}^2$, we recover the ordinary gamma functions, 
$\Gamma(\nu)$. The potential has a static remnant in the
bosonic string. Setting $u$ to zero in Eq.\ (\ref{eq:gammas}) gives 
the simple result:
\begin{equation}
V(r) = - (d-2) \cdot {{1}\over{r}} \quad ,
\label{eq:oneover}
\end{equation}
where $d$$=$$26$ is the critical spacetime dimension of the bosonic 
string. The velocity dependent corrections have an analog in the 
type ${\rm I}^{\prime}$ superstring \cite{typeI}. An analogous
static term is
present in the contribution from the Neveu-Schwarz sector \cite{typeI},
prior to cancellation by other contributions to the vacuum amplitude
\cite{polchinskibook}. It is evident from 
Eq.\ (\ref{eq:gammas}) that our result for the potential between slow 
moving sources holds for arbitrarily short distance scales lower than 
the string scale: $r^2_{\rm min.}$$\sim$$2\pi\alpha^{\prime}u$, limited 
only by the domain of validity for the double expansion in small 
velocities and short distances.

Let us compare the short distance static potential with the known 
form of the heavy quark-antiquark potential in QCD at long distances, 
a regime described by the effective dynamics of a thin flux tube 
linking the sources. The usual route from the Wilson loop expectation 
value to the static heavy quark-antiquark potential in gauge theory 
is as follows.  Consider a rectangular Wilson loop $RT$ in the limit
${{T}\over{R}}$$\to$$\infty$ with $R$ held fixed. The long legs
of the rectangle are interpreted as the proper time
world-lines of a heavy quark and antiquark, and the
loop expectation value takes the form $a(T)e^{-V(R)T}$, with 
$V(R)$ interpreted as the static quark-antiquark potential at 
fixed spatial separation $R$. $a(T)$ is some function with slower 
fall-off than an exponential. The reader may wonder why we considered 
a pair correlation function rather than extract the potential from the 
expectation value of a single rectangular loop, as is usual in gauge 
theory. The reason is Weyl invariance: the world-sheet spanning a single 
rectangular boundary loop has curvature singularities at the corners 
leading to Weyl anomalies which would render a covariant path 
integral quantization untenable. The large loop length limit hides 
this problem since the corners are pushed to $\tau$$\to$$\pm \infty$. 
The pair correlation function does not suffer from this problem. In 
particular, for any pair of coplanar nested right-angled loops 
we had a well-defined expression for the string path integral 
even {\em prior} to taking the large loop length limit.

The heavy quark-antiquark potential at long distances displays 
a confining linear plus attractive inverse power law behavior:
\begin{equation}
V(r)  =  \alpha r + \beta +
{{\gamma}\over{r}} + O(1/r^{2}) \quad .
\label{confine}
\end{equation}
$\alpha$ and $\beta$ are known to be nonuniversal constants. Of
greater interest is the universal constant, $\gamma$, first obtained
by L\"{u}scher et al using heat kernel methods in \cite{luscheret}: 
$\gamma$$=$$-(d-2)\pi/24$ in the effective theory of the Eguchi-Schild 
string \cite{eguchi}. Recall the model independent argument for the coefficient 
of the $1/r$ term \cite{luscher} (see, also, the discussion in 
\cite{effective}). Consider the quantum dynamics of a 
thin flux tube linking quark and antiquark as described by an effective 
field theory. Let $d$ be the number of degrees of freedom. Now the
fluctuations of a long thin flux tube in $d$$-$$1$ spatial dimensions are 
described by a two dimensional nonrenormalizable effective field theory 
of $d$$-$$2$ collective modes, with universal behavior that of $d$$-$$2$ 
free scalar fields, each with vacuum energy equal to $\pi/24$. The $O(1/r^2)$ 
terms that are quartic and higher order in the collective fields are irrelevant. 
The vacuum energy arises from the infinite sum of free field harmonic 
oscillators in their ground state, with an independent sum for 
each of $d$$-$$2$ degrees of freedom. Irrelevant couplings to higher 
dimensional operators can induce interactions; they determine the 
nonuniversal constants, $\alpha$, $\beta$. Not surprisingly, this long 
distance result for the potential is in agreement with our expression 
for $E_0$, the vacuum energy from each of $d$$-$$2$ free world-sheet scalars 
given in Eq.\ (\ref{eq:vacuum}).

Our computation demonstrates that there is also a universal $1/r$ static 
potential at {\em short distances}: independent of the dimensionality 
of the Dbrane worldvolume, the geometrical parameters of the loops, and
the string scale cutoff. As we 
can see from Eq.\ (\ref{eq:oneover}), the numerical coefficient at short 
distances predicted by string theory differs from L\"{u}scher's long distance 
result. This may be interpreted in the effective field theory as a 
wavefunction renormalization for the Wilson loop observable at short 
distances, an effect which {\em cannot} be determined in a field theoretic
analysis insensitive to boundary effects. Moreover, there is an 
{\em infinite} number of velocity dependent 
corrections to the $1/r$ term which are {\em also} universal. We obtained 
these corrections by a systematic double expansion in small velocities and 
short distances, conveniently expressed as a convergent power series in 
dimensionless variables, $z$$=$$r_{\rm min.}^2/r^2$, $\upz$, and $\ups$.

Our results can also be considered within the more traditional
context of phenomenological models for short distance nonperturbative
dynamics in QCD (see, for example, \cite{polyakov} and citations 
thereof). The generic backgrounds for the Liouville field with $c_m$$<$$d$ 
described in section IIIC could be of interest in this context.
We note that recent work\footnote{We would like to thank M. Eides 
for bringing this work to our attention.} on the short 
distance potential between heavy sources in QCD has examined 
modifications of the potential at short distances originating in 
nonperturbative instanton effects \cite{zakharov}.

\subsection{Minimum Distance and the Short Distance Scattering of D0branes}
\label{sec:linear}

D0branes are point-like topological defects in spacetime
present in the generic background of the bosonic string theory.
Consider the potential between two D0branes
probed in their nonrelativistic scattering \cite{bachas,kpdfs,dkps}. 
The D0branes are assumed to have fixed spatial separation $b$
in the direction $X^{d-1}$, and are in relative slow motion 
in an orthogonal direction $X^d$ with velocity $v$ \cite{bachas}.
At long distances their static interaction potential 
will take the Newtonian form. The effective potential 
at long distances is dominated by the exchange of the lowest lying states 
in the {\em closed} string spectrum. With no loss of generality, we
can obtain the potential between two D0branes as a special case of
the general expression for the scattering of two Dpbranes, $p$$<$$d$. 
We will show that the systematics of the small velocity short distance
double expansion and the value for the minimum distance probed in
the scattering, $r_{\rm min.}$, is identical to the result obtained 
above, in general agreement with previous analyses 
\cite{bachas,dkps,polchinskibook}.

Adapting the computation of the bosonic string annulus 
amplitude between static Dpbranes \cite{dnotes,polchinskibook} to the
boundary conditions pertinent to Dpbrane scattering, and 
restricting to massless modes gives:
\begin{equation}
V_{\rm Dpbrane}(r,u) = - V_p (8 \pi^2 \alpha')^{-\p1half} 
  \int_0^{\infty} dl e^{-r^2 l/2\pi\alpha' } {l}^{21-p/2} 
 \left [  {{ {\rm tanh}(u) } \over{ i {\rm Sin}(-iu)}} ( 22 + 2{\rm Cosh}(2u) )
\right ]  \quad .
\label{eq:masslessclosed}
\end{equation}
Notice that, unlike the expression for the short distance potential, a Taylor
expansion of the periodic functions in the integrand for small velocities and
long distances gives 
only $O(u^2)$ corrections to the static potential. The integration domain is 
unrestricted for small velocities. Performing the integration over $l$ gives 
the simple result:
\begin{equation}
V_{\rm Dpbrane}(r,u) = - V_p (8\pi^2 \alpha^{\prime})^{-\phalf} 
 \Gamma (\23phalf) (2\pi\alpha^{\prime})^{\23phalf} r^{p-23} \left [ 
  24 + O(u^2) \right ]  \quad .  \label{eq:coulomb}
\end{equation}
Setting $p$$=$$0$ gives the Newtonian long distance
interaction for D0branes in a $d$$=$$26$ 
dimensional spacetime.

At short distances, we will find a crossover phenomenon analogous to what was 
found in the interaction potential of a Dpbrane with a 
D${\rm p}^\prime$brane for dimensionalities, $p$$-$$p^{\prime}$$\neq$$0$ mod $4$ 
\cite{dkps}: the asymptotic long and short distance forms of the pair potential 
between Dpbranes in the bosonic string are {\em not} identical. Consider the 
expression for the effective potential due to the exchange of 
massless modes in the open string spectrum: 
\begin{equation}
V_{\rm Dpbrane}(r,u) = - V_p (8 \pi^2 \alpha')^{-\p1half} 
  \int_0^{\infty} dl e^{-r^2 l/2\pi\alpha' } {l}^{-\p1half} 
 \left [  {{ {\rm tanh}(u) e^{-u^2l/\pi} } \over{ {\rm Sin}(ul)}} 
( 22 + 2{\rm Cos}(2ul) )
\right ]  \quad .
\label{eq:masslessopen}
\end{equation}
The small velocity short distance double expansion can be performed as explained
in the previous section. The result is the expression:
\begin{eqnarray}
V_{\rm Dpbrane}(r,u) =&&
 - V_p (8\pi^2\alpha^{\prime})^{-\p1half} 
 {\rm tanh}(u)/u [  
 24 {\mathbf{\gamma}} (- \p1half , (1+\upz)/z) 
 \cdot ({{r^2}\over{2\pi\alpha^{\prime}}})^{(p+1)/2}(1+\upz)^{(p+1)/2} 
\nonumber \\
&& \quad\quad 
+ O(z^2,\upz,\ups) ] \quad ,
\label{eq:linear}
\end{eqnarray}
where the velocity dependent corrections are obtained in a systematic
expansion in the same dimensionless variables, $z^2$, $\upz$, and $\ups$, 
defined above. For $r^2$$>>$$r_{\rm min.}^2$, $z$$\to$$0$, we 
recover the gamma functions, $\Gamma(-\nu)$. Recall that gamma functions 
with negative argument can be defined by iterating the well-known 
identity, $-\nu \Gamma(-\nu)$$=$$\Gamma(-\nu+1)$. We note that the short 
distance static potential between D0branes is linear, and {\em repulsive}:  
\begin{equation}
V_{\rm D0brane}(r) = - (d-2) {{r}\over{2 \pi\alpha^{\prime}}} + 
 O(z^2, \upz,\ups) \quad .
\label{eq:dzero}
\end{equation}
This result holds in a self-consistent small velocity short distance 
approximation with corrections 
of $O(z^2,\upz,\ups)$. It is valid for distances in the range, 
$2\pi\alpha^{\prime}u$$<<$$r^2$$<<$$2\pi\alpha^{\prime}$, and velocities in the 
range, $u$$<<$$u_+$, where the upper bound is estimated as described in section
IVA.

The static potential between D0branes corresponds to the vacuum energy in a 
background of open string theory with constant electromagnetic potential, $A^{d-1}$, 
but with vanishing electric field strength: $E^{d-1}$$=$$\partial_0 A^{d-1}$$=$$0$ 
\cite{dbrane,bachas,polchinskibook}. The potential is a measure of the shift in
the vacuum energy relative to that in the background with no Dbrane sources.

\section{Conclusions}
\label{sec:conc}

Our computation in open and closed string theory is performed at weak
coupling in flat spacetime backgrounds and in the critical spacetime 
dimension. There is a supersymmetric analog to this result which will be 
explored in forthcoming work \cite{typeI}. We have 
demonstrated the validity of the double expansion in small velocities and short
distances down to a minimum distance, $r^2_{\rm min.}$$=$$2\pi\alpha^{\prime}u$,
in general agreement with previous estimates \cite{shenker,bachas,dkps,polchinskibook}.
Thus, String/M theory predicts an {\em infinite} number
of velocity dependent corrections to the potential between two heavy sources 
in relative slow motion in a gauge theory, the numerical coefficients of 
which are predicted by a systematic expansion. We are not aware of a comparable 
theoretical analysis which reliably probes this regime of QCD. We note that
there is no evidence of non-analytic behavior in the potential between sources
at short distances, suggesting that the phase transition at short distances 
previously found in \cite{alvarezstatic,durhuus} is a large $d$ artifact.

The numerical difference between the coefficient of the $1/r$ term we have
found in the static potential at short distance and that given by L\"{u}scher's 
effective field theory analysis for the QCD flux tube valid at long distance 
scales deserves explanation. We should note that the flux tube picture is 
inherently a long distance concept whose predictions cannot be naively 
extrapolated to short distances. Consider the $d=26$ dimensional bosonic string. 
At both long and short distances there is a proportionality factor in the $1/r$
potential which equals $d-2$, the number of transverse massless gluon modes. The 
short distance static potential between heavy point sources in the gauge theory 
is a measure of fluctuations in the vacuum energy density. It would be gratifying
if one could exploit the direct calculation of the short distance potential 
from string theory given in this paper to explore nonperturbative 
physics associated with the QCD vacuum at short distances, a subject rich
in conjecture and in open questions \cite{polyakov}. This deserves further study.

We conclude by noting that the characteristic D0brane velocity in the 
supersymmetric theory is of order $u$$\simeq$$g_s^{2/3}$ \cite{kpdfs}, 
which implies a minimum distance
$r_{\rm min.}^2$$\simeq$$\alpha^{\prime}g_s^{2/3}$. It is interesting to 
note that, at large $N$, the shortest distances that can be probed in 
the small velocity short distance approximation are pushed down to 
$r_{\rm min.}^2$$=$$g_{\rm eff.}^{2/3}/N$ \cite{dkps}, where 
$g_{\rm eff.}$$=$$g_s N$. In the introduction, we emphasized the 
importance of taking a large $N$ limit in order to keep gravitational 
corrections to amplitudes in the open string sector suppressed at
{\em long} distances; this was also an essential observation underlying 
Maldacena's conjecture \cite{malda}. We see now that taking the large N 
limit extends the regime of weakly coupled open and closed string theory 
{\em both} in the directions of longer {\em and} of shorter distance scales. 
We believe it would be of great interest to develop a systematic 
formulation of the large $N$ limit of open and closed string theory.

\vspace{0.3in}
\noindent{\bf Acknowledgments}

S.C. would like to thank M. Douglas, N. Drukker, M. Eides, S. Giddings, 
D. Gross, A. Hashimoto, D. Kabat, D. Kutasov, G. Moore, J. Polchinski, 
S.-J. Rey, S. Shenker, and H. Tye for their comments, and the Institute for 
Theoretical Physics and the Aspen Center for Physics for their hospitality. 
This work is supported in part by the National Science Foundation grant 
NSF-PHY-97-22394.



\end{document}